\documentclass[aps,twocolumn,groupedaddress]{revtex4-1}

\usepackage{graphicx}% Include figure files
\usepackage{color}

\begin{document}

\title{Slow ground state molecules from matrix isolation sublimation}

\author{A.~N.~Oliveira$^{* 1,2}$, R.~L.~Sacramento$^1$,  B.~X.~Alves$^1$, B.~A.~Silva$^1$, W.~Wolff$^1$ and
C.~L.~Cesar$^1$}
\affiliation{$^1$Instituto de F\'{\i}sica, Universidade Federal do Rio de
Janeiro,  Caixa Postal 68528,
21941-972 Rio de Janeiro, RJ, Brazil\\
$^2$INMETRO, Av. Nossa Senhora das Gra\c{c}as, 50\\
Duque de Caxias - RJ, Brazil\\
$^*$nunes@if.ufrj.br
}

\date{}

\date{\today}

\begin{abstract}
We describe the generation and properties of a cryogenic beam  of  $^7$Li$_2$ dimers from sublimation of a neon matrix where lithium atoms have been implanted via laser ablation of solid precursors of metallic lithium or lithium hydride (LiH). Different sublimation regimes lead to pulsed molecular beams with different temperatures, densities and forward velocities. With laser absorption spectroscopy these parameters were measured using the molecular $^7$Li$_2$ (R) transitions A$^1\Sigma_u^+(v'=4,J'=J''+1)\leftarrow $X$^1\Sigma_g^+
(v''=0,J''=0,1,3)$. In a typical regime, sublimating a matrix at 16~K, translational temperatures of 6--8~K  with a drift velocity of 130~m$\,$s$^{-1}$ in a free expanding pulsed beam with molecular density of 10$^9$~cm$^{-3}$, averaged along the laser axis, were observed. Rotational temperatures around 5--7~K were obtained. In recent experiments we were able to monitor the atomic Li signal -- in the D2 line -- concomitantly with the molecular signal in order to compare them as a function of the number of ablation pulses. Based on the data and a simple model, we discuss the possibility that a fraction of these molecules are being formed in the matrix, by mating atoms from different ablation pulses,  which would open up the way to formation of other more interesting and difficult molecules to be studied at low temperatures. Such a source of cryogenic molecules have possible applications encompassing fundamental physics tests, quantum information studies, cold collisions, chemistry, and trapping.
\end{abstract}

%\submitto{\jpb}
\maketitle

\section{INTRODUCTION}

The production of cold samples or beams of molecules is an ongoing research field since it has many
important potential applications as discussed in \cite{ColdMolsReviewNJP2009}. Among the interesting applications   there is the prospective use in quantum information and computation~\cite{DeMille2002, DeMille2006} and basic physics tests, such as the search for a permanent electric dipole moment of the electron~\cite{HindsEDM2011, Baron2014} with its corresponding parity and time reversal symmetries violations.  Some molecular species are complex to make in an environment compatible with low temperatures~\cite{Walker93}, requiring the  use of hot ovens. Low velocities and particularly low temperatures are important for long or precise interaction times -- with lasers or microwave -- in high precision experiments. As an illustration, in precision experiments, the work in \cite{HindsEDM2011} used a supersonic beam of YbF with a forward velocity of 590~m$\,$s$^{-1}$  while the work in \cite{Baron2014}  used a buffer gas precooled beam of ThO with forward velocity of 200~m$\,$s$^{-1}$ and temperatures of $\sim$10~K.

The study of ultracold collisions in magneto-optical traps (MOT) and photo-association experiments have produced large amounts of ultracold
molecules, but only with a few atomic species which are amenable to be trapped in a MOT~\cite{ColdMolsReviewNJP2009,Marcassa2004, Weidemuller2012}. A whole field of cold chemistry will benefit from the availability of different species of cold molecules. The easier the technique to produce cold molecules, trapped or as beams, the more applications will be found, from sensors, to basic physics and chemistry.

In the field of Matrix Isolation Spectroscopy (MIS), molecules have been studied inside the
inert-gas solid  matrix \cite{LeeScience2006,FajardoH2,Fajardoexp,Cradock75,Ball88}, including
rotating molecules in a solid H$_2$ matrix \cite{LeeScience2006}. Optical absorptions due to
molecular lines have been observed, albeit with large frequency broadening and shifts from their
transition values in free space. Fajardo and coworkers \cite{LeeScience2006,FajardoH2,Fajardoexp} report an annealing treatment to the matrix resulting in better resolved molecular lines.

In the present work we expand our research program on Matrix Isolation Sublimation (MISu) -- where
intense cryogenic atomic beams of Cr and Li originating from Ne or H$_2$
matrices have been studied \cite{laserJCP2012,laserJCP2011,laserCJP2009,laserPRA2007}  -- to demonstrate a cryogenic beam of $^7$Li$_2$ dimers sublimated from a Ne matrix. The transitions in $^7$Li$_2$ dimers are  accessible by the experimental setup and diode lasers built in the laboratory and also allow for a simultaneous comparison of the molecular signal to the atomic signal which has been extensively studied. Thus, the choice of $^7$Li$_2$ dimers, while circumstantial, serves as a first demonstration and evidence of the possibilities of the MISu technique. Most importantly, it enabled us to start probing the nature of the molecular formation process and answer questions such as: would the molecules be released in a variety of rovibrational states due to possible strains in the matrix? And, what is its temperature and velocity distribution? The pulsed molecular beam resulting from the MISu technique appears attractive in terms of density, temperatures and forward velocities for many applications. One of its main advantage may be the easiness by which one could form molecules which are not accessible by the other techniques.

There are three non-excluding possibilities for formation of the molecules detected in our experimental procedure: (i) they are implanted as molecules from the laser ablation process~\cite{Milosevic} ; (ii) they are formed in the matrix when the implanted atoms fall adjacent to each other; (iii) they are formed during the matrix sublimation, in flight. While all three processes could be involved, process (ii) is clearly very attractive as it would  open up the way to produce interesting heteronuclear molecules that are difficult to produce in compatibility with high precision measurements, with low temperatures and forward velocities. We discuss, in view of the data and some simple models, these three processes pointing to evidence that a relevant fraction of the molecules we detect should be formed by the second process.

This first molecular MISu study leave interesting open questions: - what is the percentage of molecules formed in the matrix as opposed to implanted directly from ablation, and how does that depend on the size, or composition, of the molecule? - what is the limit on the number of molecules formed (we did not reach beyond 5\% of laser absorption)? - will softer molecules suffer from strains in the matrix  leaving a low population in the ground rovibrational state upon sublimation? - what will be the effect of a controlled annealing of the matrix to the number of molecules?

\section{EXPERIMENTAL SETUP}

\begin{center}
\begin{figure}[ht]
\includegraphics[width=2.0 in]{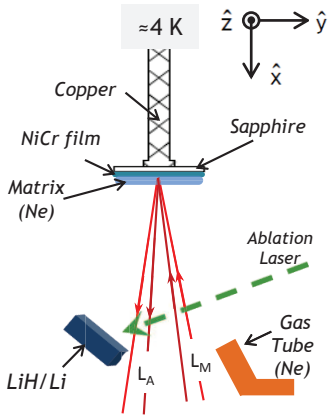}
\caption{Schematics of the experimental apparatus showing the sapphire substrate, the NiCr film resistor and the deposited matrix of Ne which comes from the gas tube. Li atoms, and perhaps molecules, are implanted via laser ablation (shown in dashed green) on a solid Li or LiH precursor. The spectroscopy laser beams (shown in red and identified as L$_{\rm A}$ and L$_{\rm M}$), propagate  along the direction of the sublimation plume expansion and are also used to monitor the matrix film thickness.}
\label{Fig:experimentalsetup}
\end{figure}
\end{center}

The  experimental setup has been described in detail elsewhere~\cite{laserJCP2012,laserJCP2011} and a diagram is shown in figure \ref{Fig:experimentalsetup}. The system consists of
a closed-cycle cryostat, with optical access, that reaches 3~K at its cold plate with 1~W cooling
power at 4~K. A sapphire substrate, with a deposited resistive film of NiCr, is thermally connected,
through a chosen thermal link, to the cold plate. This configuration permits large and fast variations of the sapphire substrate temperature without affecting the cryostat temperature. The Ne gas that forms the matrix is delivered through a 2~mm inner diameter tube whose exit points towards the sapphire substrate. The flow rate of the gas is controlled by the pressure in a reservoir followed by a high impedance line, achieving 1--10~mmol/hr.
The growth of the Ne matrix film is monitored by two CW laser beams -- called ``atomic'' (LA) and ``molecular'' (LM) lasers --  through the interference fringes (with 10--12\% visibility) off the etalon formed by the Ne film together with the $\sim$40\% reflectivity of the NiCr resistive film on the sapphire substrate. These  beams come from the bottom upwards, with slightly different angle, reflect off the Ne film and sapphire mirror at near normal incidence, and are monitored by photodiodes as indicated in figure \ref{Fig:experimentalsetup}. A typical time for growing a $\lambda/2$ thick Ne layer in the data here presented is 40~s, but it can be reduced to less than 1~s.

The lasers are tuned to the atomic D2 transition in $^7$Li (at 670.776 nm -- air wavelength is used throughout the paper) and the molecular $^7$Li$_2$ (R) transitions A$^1\Sigma_u^+(v'=4,J'=J''+1)\leftarrow $X$^1\Sigma_g^+ (v''=0,J''=0,1,3)$ (at 665.948 nm, 665.924 nm, 665.927 nm, respectively). The Franck-Condon factor for this molecular transition \cite{Barakat} is 0.158. The transitions energies have been calculated by
taking \cite{LeRoy,Coxon} $T_e$(A-X) = 14068.185~cm$^{-1}$, G$_0$= 175.029~cm$^{-1}$ and B$_0$=0.669~cm$^{-1}$ for the X$^1\Sigma_g^+$ state and G$_4$=1117.918~cm$^{-1}$  and B$_4$=0.4732~cm$^{-1}$ for the A$^1\Sigma_u^+$ state and checked against the data of D. Hsu \cite{Hsu}. The laser beams have a diameter around 4~mm and powers of  tens of microwatts to avoid much saturation.

The experiment proceeds in a sequence as follows. Allowing Ne to flow into the cryostat a desired layer of Ne solid matrix is formed. Then, a pulsed laser, at 1~Hz  repetition rate and up to 25~mJ of energy per pulse, promotes ablation of the solid precursor (metallic lithium or LiH), implanting atoms, and possibly molecules, into the matrix.  The laser is focused onto the solid precursor by a 75~mm lens. The number of ablation pulses is varied from 1 to 300. Special care should taken to avoid Ne matrix loss -- which could be observed by the laser interference fringes -- due to the temperature increase when absorbing ablation energy. For some experimental conditions we had to lower substantially the energy of the pulsed laser to avoid this matrix loss. During ablation, very large absorptions on the atomic spectral line are seen, while no absorption on the molecular line is observed. This is expected from the high temperatures achieved during the laser ablation phase which would diminish the population of the particular molecular ground-state seen by the laser. For some data -- in the case of thinner matrices ($13-60~{\rm nm}\ll \lambda/2$) -- the cycle above is repeated up to 5 times.

After the matrix is completed, a heat pulse is applied to the sapphire NiCr resistor causing the
matrix to sublimate into vacuum while the absorption spectra is recorded by scanning the lasers' frequencies. The sublimation time depends on the heat pulse energy and was typically between 60--240~$\mu$s. At longer sublimation times, we reach the limit of sensitivity of our detection for molecules, while with atoms we have seen signals even at sublimation times of seconds, reaching a quasi-CW operation. During this period the sample expands and its center-of-mass moves in $\mathbf{\hat{x}}$ direction (see figure \ref{Fig:experimentalsetup}).  This flight direction leads to a splitting of the resonance lines due to a positive and a negative Doppler effect, since the laser beams are both co-propagating and counter-propagating with the sublimation plume. This allows for a direct measure of the sample's forward velocity. The Doppler profile enables the measurement of the sample's longitudinal temperature (in other studies we have investigated the transverse temperature as well \cite{laserJCP2012,laserJCP2011}). The lasers frequency scans are monitored using two 1~GHz free spectral range Fabry-Perot interferometers.

The whole experiment can be repeated in time scales of one minute, but with a lower gas impedance this time can be brought down to a couple of seconds. The data analysis starts by finding the best fit for the atomic and molecular Fabry-Perot signals -- disregarding the small peaks (modes other then the fundamental) -- thus allowing a plot of photodiode signals as a function of frequency. Then, we can subtract the lasers amplitude modulation, by fitting it at longer times, in absence of atomic or molecular absorption, and extending to the earlier times. The interference fringe from the Ne matrix sublimation at early times can also be subtracted by a complex thermal model \cite{laserJCP2012} or by a simple linearization near the absorption peaks. Finally, the spectrum is adjusted by a double gaussian yielding several parameters, such as: peak optical density, forward velocity and temperature of the molecular and the atomic beams.

\section{EXPERIMENTAL DATA AND DISCUSSION}

\begin{center}
\begin{figure}[ht]
\includegraphics[width=3.0 in]{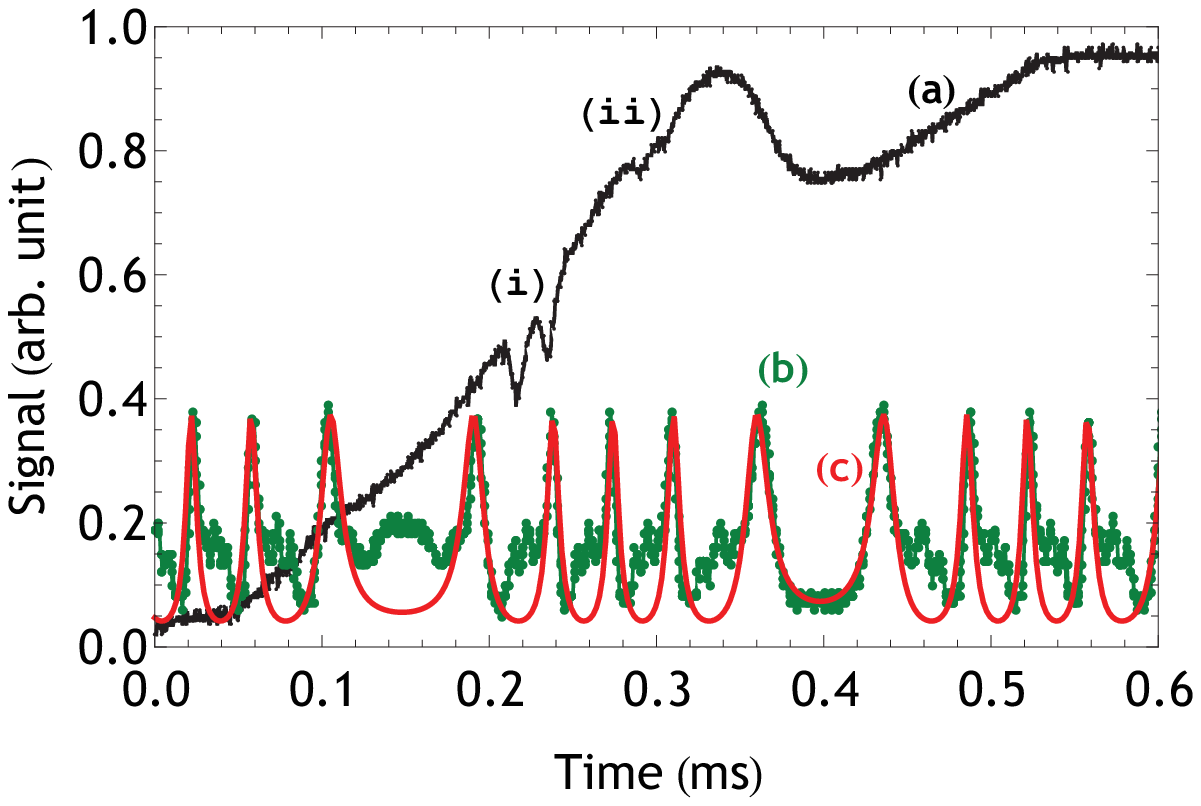}
\caption{Raw data for laser scan around 665.924 -- 665.927 nm, showing the laser transmission (a), the Fabry-Perot signal (b) with a fit (c). The laser transmission signal is offset in the figure. It has a residual amplitude modulation due to the laser PZT scan, and it is affected by the Ne matrix sublimation, displaying a fraction of an interference fringe. In a single scan we are able to observe two rotational levels: J''=1 (i) and J''=3 (ii). See text for more details.}
\label{Fig:rawspecj13}
\end{figure}
\end{center}

Following the experimental procedure, a typical data  is presented in figure \ref{Fig:rawspecj13}. The laser power transmitted through the sample upwards and backwards to the photodiode is also sensitive to the Ne matrix sublimation, which is evident in the large amplitude variation in the first 0.2~ms. In this figure, a single scan captures two transition lines  A$^1\Sigma_u^+(v'=4,J'=2)\leftarrow $X$^1\Sigma_g^+(v''=0,J''=1)$ and A$^1\Sigma_u^+(v'=4,J'=4)\leftarrow $X$^1\Sigma_g^+(v''=0,J''=3)$ (at 665.924 nm and 665.927 nm, respectively).

\begin{center}
\begin{figure}[ht]
\includegraphics[width=3.0 in]{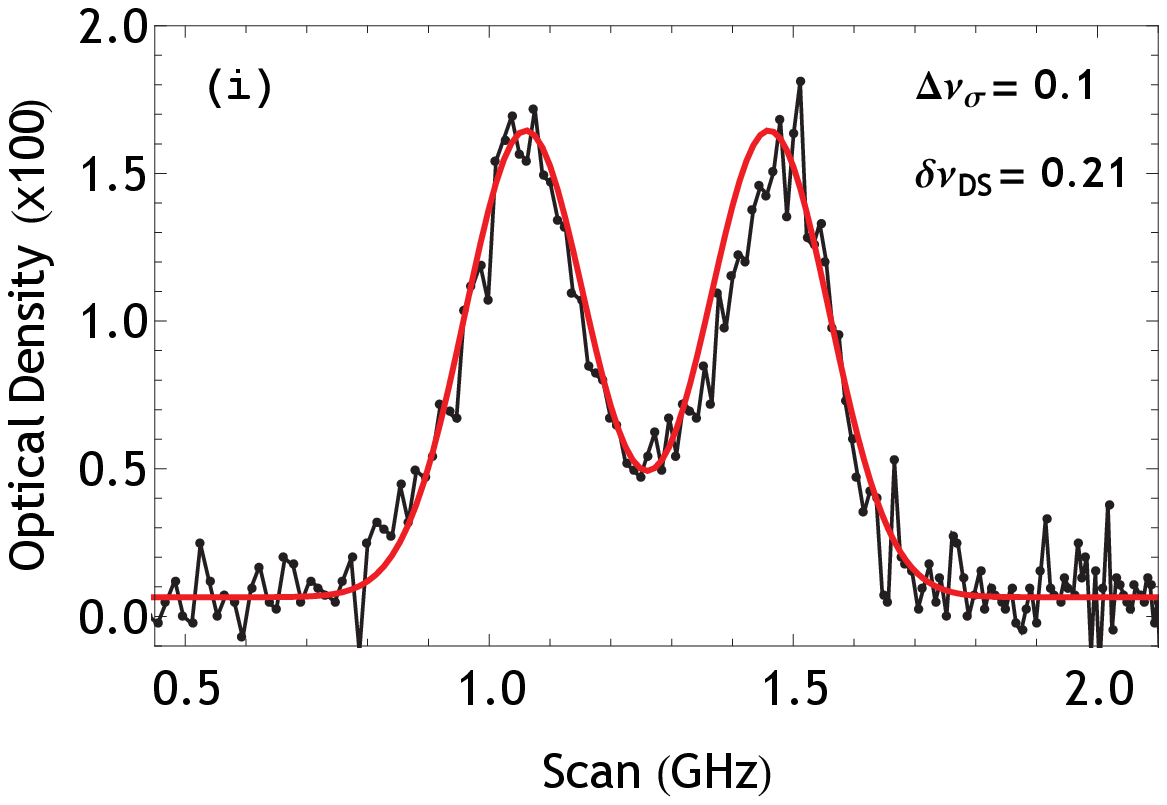}
\includegraphics[width=3.0 in]{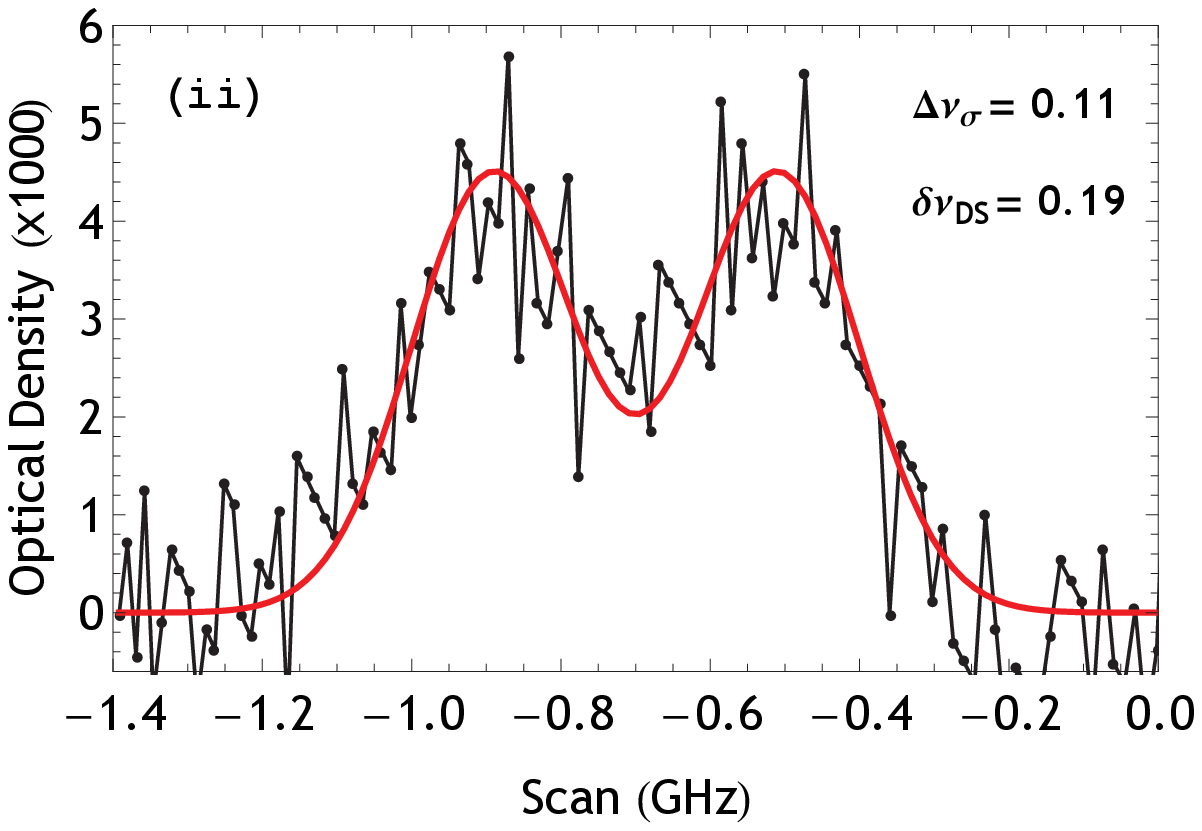}
\caption{Treated data for the peaks (i) and (ii) from figure \ref{Fig:rawspecj13}  showing visual fits  and their parameters~\cite{laserJCP2012}. Taking into account a similar set of data and using mainly the information on the strongest signal we find the parameters to be compatible with the following values: forward velocity $v_{\rm CM} \approx$ 130 -- 140~m$\,$s$^{-1}$ and longitudinal translational temperature $T_{\rm long}  \approx 6-8~$K. From the sublimation time we infer a sapphire temperature $T_S \approx 16~$K. From the optical densities one can obtain the molecular density along the laser path.}
\label{Fig:specj13}
\end{figure}
\end{center}

The distance between peaks in each doublet in figure \ref{Fig:rawspecj13} -- (i) and (ii) -- accounts for the forward (center--of--mass) velocity, while the gaussian profile, once fitted in frequency domain, measures the longitudinal translational temperature as shown in figure \ref{Fig:specj13}. We fit the gaussian standard deviation $\Delta\nu_\sigma=\sqrt{k_B T_{\rm long}/M_{\rm Li_2}}/\lambda$, and the Doppler shift between the two components $\delta\nu_{DS} = v_{CM}/\lambda$, obtaining the forward velocity $v_{\rm CM}$ and the longitudinal temperature $T_{\rm long}$, where $M_{\rm Li_2}$ is the mass of the lithium dimmer, $k_B$ is Boltzmann's constant, and $\lambda$ is the wavelength.

\begin{center}
\begin{figure}[ht]
\includegraphics[width=3.5 in]{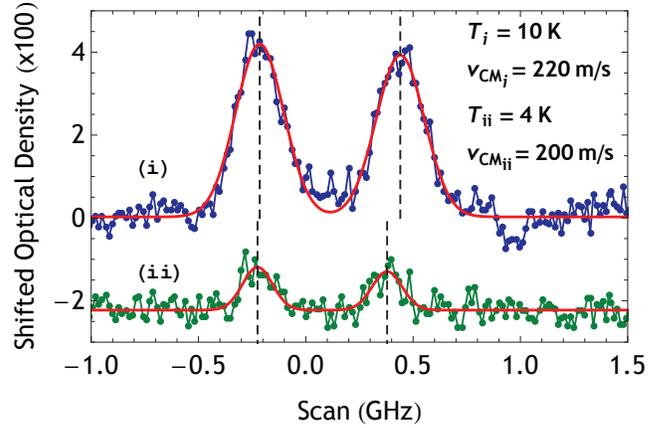}
\caption{ A spectrum at 665.949 nm at two subsequent scans during the same sublimation. In the second one we can see a lower density and also lower velocity and temperature.}
\label{Fig:spec_mol_j14}
\end{figure}
\end{center}

A typical signal in the A$^1\Sigma_u^+(v'=4,J'=1)\leftarrow $X$^1\Sigma_g^+(v''=0,J''=0)$ transition at 665.949~nm is shown in figure \ref{Fig:spec_mol_j14}.

In order to reach a reasonable signal-to-noise ratio in the molecular signal we resorted to fast sublimations. In most of the experiments, the matrix was sublimated within 60 to 240~$\mu$s. The laser resonance condition was often met at a later time, after the full matrix had sublimated. As can be easily imagined and corroborated with a Monte-Carlo simulation, there is a dynamical reduction of temperature for times after the full matrix has sublimated, both in the molecular and the atomic samples. Essentially, the fast atoms or molecules go quickly outside the laser beam and towards the walls where they stick, while the slower species remain longer within the laser beam. This effect can be avoided with atoms, where the large signal-to-noire ratio allows for a slow sublimation with the laser interacting with a freshly released sample, giving a more instantaneous measurement. In this time delayed process we can observe  temperatures of the atomic beam down to 1--2~K.

Most of our data was obtained using a cell of 75~mm diameter and a distance between the sapphire substrate and the bottom window of 65~mm. With a forward velocity of $\sim$130~m$\,$s$^{-1}$, the sample's center--of--mass would reach the window at 500~$\mu$s (longer than the typical time at which we get our signals). Moreover, as the atoms or molecules move away from the sapphire they have smaller probability of passing through the laser beam, this being true for all the directions. We did a quick test on the atomic signal with an experimental cell 4 times larger in all the dimensions and got identical results. So, our quoted results are not influenced with collisions to the surfaces. Only in special experiments, after many cycles and having released a large quantity of Ne in the cell and with large atomic signals, we can observe very small ``revival'' signals from Li atoms reflecting off the Ne coated cell windows.

The longitudinal translational temperature measured for the molecules is in accordance to our predicted 1--d or 3--d thermalization models~\cite{laserJCP2012}. For example, for the data in figure \ref{Fig:rawspecj13} (and figure \ref{Fig:specj13}), where the sublimation temperature is $T_S\sim16$~K and the molecular signal is seen during the sublimation, the 1--d thermalization model would predict a translational longitudinal temperature $T_{1d}=T_S (4-\pi)/2 \sim 7$~K which is in  agreement with the measured temperature.

On the other hand, the forward velocity seems higher than our previous models would indicate. We can deduce the substrate temperature from the sublimation time. For the data in figure \ref{Fig:rawspecj13} (and figure \ref{Fig:specj13}) we obtain a substrate temperature of 16~K, and we measured forward velocities around $\sim$130~m$\,$s$^{-1}$, which is more than the expected full entrained value, which would be Ne velocity at 104~m$\,$s$^{-1}$, and still higher than that expected from Li$_2$ being released without entrainment, which would be around 124~m$\,$s$^{-1}$. We do not know where this discrepancy come from.

From the data presented in figure \ref{Fig:specj13} we can obtain the rotational temperature ($T_{\rm rot}$) from the relative absorption line strength for different transitions by considering not just the degeneracy of the initial state, $g(J'')=2J''+1$, but also taking into account the final J' value by using the average \cite{Herzberg} $\left< J'' \right> \rightarrow (J'+J'')/2$. Herzberg's \cite{Herzberg} equation (III.169) reads:
\begin{eqnarray*}
I_{\rm abs}=\frac{C_{\rm abs} v''}{Q_r}(J'+J''+1) \exp \left[ -\frac{B_{v''} J''(J''+1)}{k_{\rm B} T_{\rm rot}}\right] ,
\label{Eq:Trot}
\end{eqnarray*}
where $C_{\rm abs} v''/Q_r$ does not depend on $J''$ or $J'$, $B_{v''} J''(J''+1)$ gives the energy of
the $(v'',J'')$ level, and $k_{\rm B}$ is the Boltzmann constant. The relative strength of these two lines point to a rotational temperature $T_{\rm rot} \approx 4 - 5~$K, even slightly below the translational temperature. This temperature range is compatible if we also include the  A$^1\Sigma_u^+(v'=4,J'=1)\leftarrow $X$^1\Sigma_g^+(v''=0,J''=0)$ (at 665.948 nm) transition signal, with its typical higher optical density maximum, from different experimental runs in the same day and conditions. But since the experiment did not have a very good reproducibility from run to run, we avoid extending the analysis, obtaining more precise values,  to other lines which were not accessible in a single run.

The fact that the rotational temperature seems lower than the translational temperature suggests that it would not be decreasing from a very high value towards equilibration with the translational temperature, which would probably be the case if the molecules were being formed during the sublimation: in this case, due to the larger phase-space of recombination, a wide variety of rovibrational states were expected to be populated and the rotational temperature would be high.

The peak molecular density can be estimated, using equation 8.7-10 of \cite{YarivBook}, from a peak absorption
of 5\% at a linewidth of 0.2~GHz and taking into account the Franck-Condon factor. The decay rate of the molecular electronic state is approximately the same as the atomic state~\cite{Chung1999}. If we average over the whole path length of the laser (6.5~cm) we get an under-estimated value of over 10$^9$~cm$^{-3}$.

\begin{center}
\begin{figure}[ht]
\includegraphics[width=3.0 in]{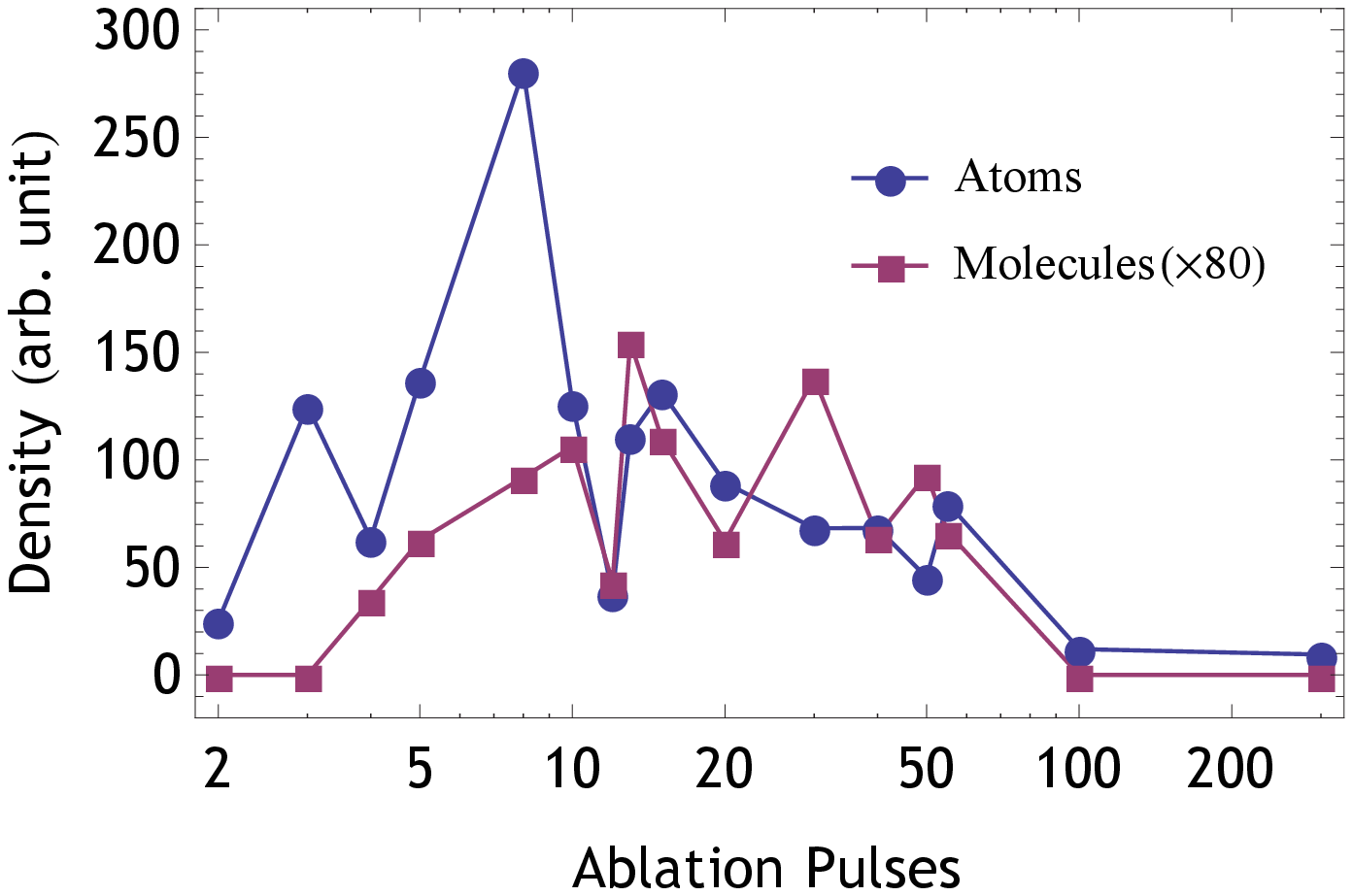}
\includegraphics[width=3.0 in]{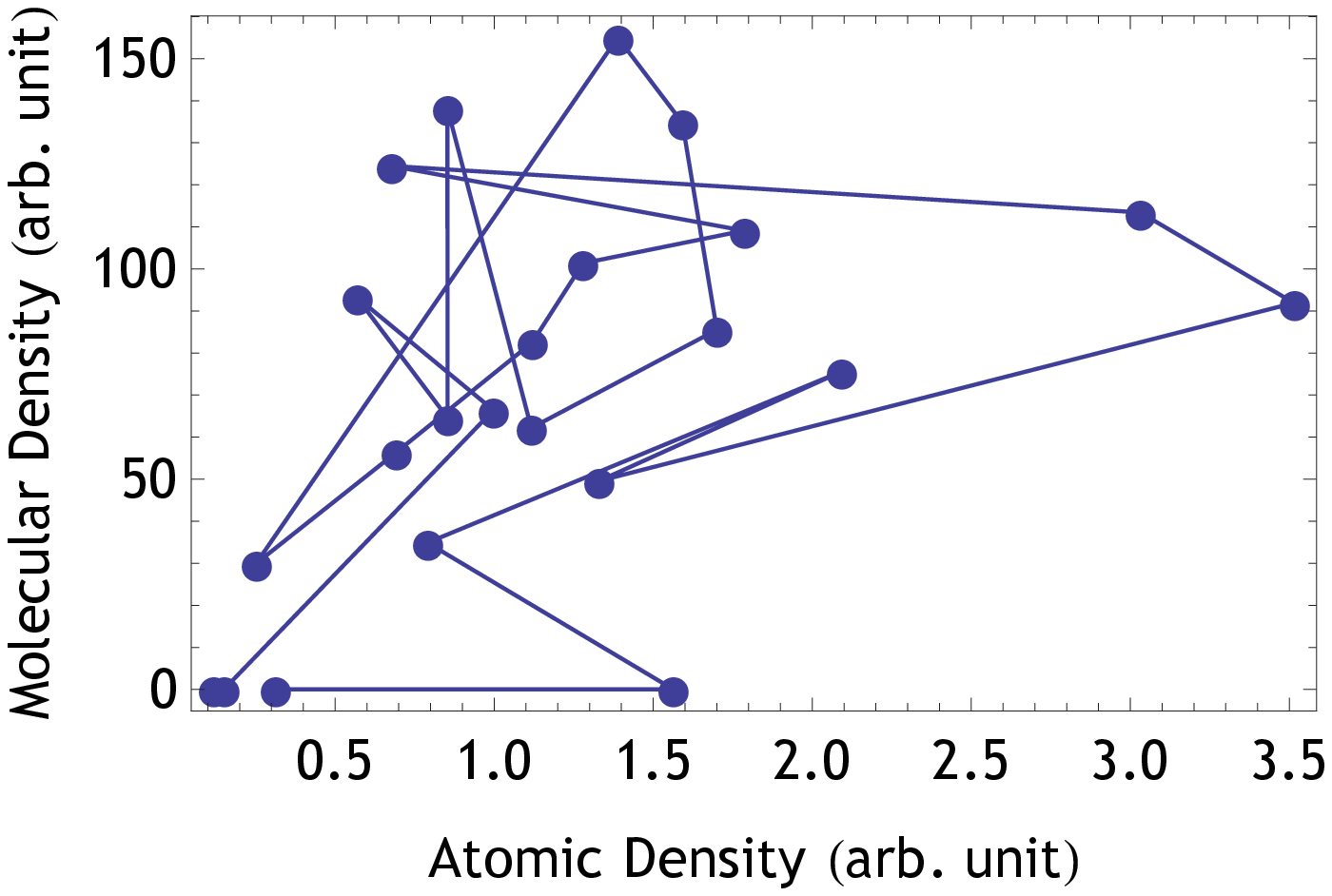}
\caption{Dependency of atomic and molecular optical densities as a function of implanted atoms (ablation pulse) in the top, showing a clear saturation. In the bottom, the molecular signal is plotted as a function of the atomic signal parametrized by the number of ablation pulses. The top graph is made with averaged signals while the bottom one employs no average. The solid lines are included just to guide the eyes. See text for discussion.}
\label{Fig:atoms_molecules_vs_ablationExp}
\end{figure}
\end{center}

We were also interested in testing the dependency of the molecular signal as a function of the number of ablation pulses. But it was soon realized that there were large variations in the signal due to the ablation process burning a hole in the solid precursor. By also monitoring the atomic signal, in the same runs, we do get some qualitative information on these dependencies. If the molecules were being formed in the evaporation of the area surround the ablation pulse \cite{Milosevic2} and the process was repetitive, we would expect both the atomic and molecular signals to increase with the number of ablation pulses and there would be a linear correlation between these absorption signals. If the molecules are formed by collisions in the ablation plume, and if this is a repetitive process one would still observe a linear growth of the atomic and molecular signals with the number of ablation pulses, while from pulse to pulse, the dependence of molecules number to atoms number would show a more rich dependence, as it involves the collision of two Li atoms with a third body that could be H, Li, or LiH.  Rather than seeing a linear, or perhaps monotonic, growth of the signals with the number of ablation pulses, we do see a clear saturation towards the disappearance of both signals. If the molecules are being formed in the matrix, we would expect an initial quadratic dependence on the number of molecules to number of atoms and a saturation of the signals as the number of ablation pulses is increased as it is  discussed in the next section. The lack of good reproducibility and the noise in the data did not allow us to to quantify the linear and quadratic dependencies between the molecular and atomic signals. The data on the dependence of the signals with the number of ablation pulses, as well as the correlation between molecular and atomic signal is shown  in figure \ref{Fig:atoms_molecules_vs_ablationExp}. This data can be better understood within the discussion in the next section.

In some series we started off saturated and saw only a decrease of the signal as a function of the number of ablation pulses. In these series we took special care to monitor the matrix thickness so that a possible matrix loss would not be accountable for this loss of signal. In some conditions we do notice the matrix evaporating away as we deposit more and more ablation pulses. In those cases the ablation laser energy was lowered and a new series attempted, until this problem disappeared.  It is clear that there is a saturation process in the matrix that leads to a decrease of both the atoms and molecules signals as the number of ablation pulses increases. We discuss a model regarding this issue in the next section.

\section{Models for molecular formation}

Lithium dimers have been observed directly from laser ablation of a metallic Li precursor using the technique of cavity ring-down spectroscopy~\cite{Milosevic}. Experiments from the same group~\cite{Milosevic2}  have detected heteronuclear molecules with alloys. Another group~\cite{DoyleCaH} have produced CaH and other molecules from laser ablation of a solid that contains the molecules, sometimes in a more complex form, such as CaH$_2$ for the case of CaH. So, clearly, dimer formation does occur during the ablation phase.  We interpret that the work in \cite{Milosevic2} would suggest that lithium dimers could be formed even from the ablation of solid LiH.

Molecular formation during the ablation process can occur via two possibilities: the dimers being formed by evaporation of the surrounding hot spot, or  being recombined in the ablation plume. In both cases the total number of dimers produced would be proportional to the number of ablation pulses, if the ablation results would be stable. If some dimers are recombining in the ablation plume, since it is a more than a one-body body process their number would show higher order dependency on the number of atoms produced in each ablation pulse, but the total number would still scale linearly with the number of ablation pulses, if they were similar.

In a regime of very low dilution of the atomic and molecular species in the Ne matrix, one would expect the atomic and molecular signal to grow indefinitely with the number of ablation pulses. Clearly this is not physical and has to fail at higher dilutions. The saturation of this process leads us to consider the possibility of forming molecules in the matrix, by atoms being implanted side-by-side.

\begin{center}
\begin{figure}[ht]
\includegraphics[width=3.0 in]{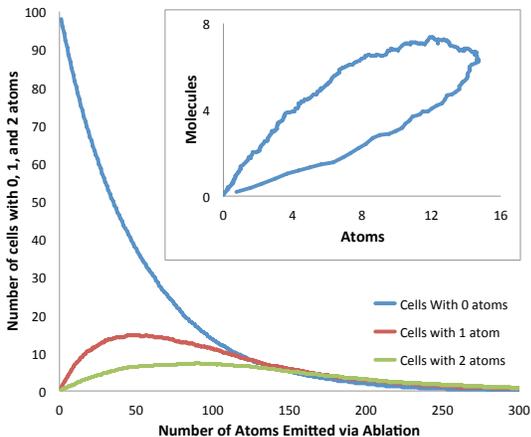}
\caption{A Monte-Carlo simulation of $n_a$ atoms thrown randomly at $n_b$ boxes or sites. In this
case, $n_b$=100, and $n_a$ varies from 0 to 500. The number of sites with 0 atoms decrease
monotonically while the ones with 1, 2 or more atoms grow to a maximum and then decrease with the
number of implanted atoms. In a very simple model we identify sites with 2 atoms as having formed
dimers, or trimers with 3 atoms, and so on. The process of forming molecules in the matrix from this
simple model would grow quadratically initially -- after all it is a 2-body process -- and would
eventually saturate and decrease. }
\label{Fig:atomsandboxes}
\end{figure}
\end{center}

We have simulated a very simple model where we consider the Ne matrix as a set of boxes, where atoms and molecules could be trapped. For a simple qualitative picture one could consider one box for each unit cell in the solid. A Monte-Carlo simulation throws atoms randomly at the boxes and one can then count the number of empty boxes, or boxes with 1 atom, or with 2 atoms, and so on. We consider  2 atoms in a box as a dimer formation, and 3 atoms as a trimer. The simulation counts the number of empty boxes, with 1 Li atom, or with 1 Li$_2$ dimer as a function of the number of atoms evaporated. We can also take into account molecules emitted directly by evaporation from the ablation by considering that a percentage of the ablation happens in molecular form. The result of one such simulation is shown in figure \ref{Fig:atomsandboxes}, where we consider ablation from LiH emitting one hydrogen atom for each lithium atom, and one H$_2$ molecule for each Li$_2$, but these molecules being emitted with a 20\% probability. The inset show the molecular population in the boxes as a function of the atomic population. The higher the probability of emitting molecules, the more linear is the growth of the molecular population in the low dilution regime. As this probability decreases, the initial growth of the molecular signal becomes fully quadratic in the number of atoms, and the inset would show a more round shape(keeping the choice of scale where the maxima -- atomic and molecular -- fill the scales).

With the formation of molecules in the matrix, the atomic signal (boxes with 1 atom) grows with the number of ablations, reaches a maximum and start decreasing. The molecular signal, will reach a maximum after the atomic passed through its maximum, and both signals go to zero in the limit of very high dilution (atoms ablated compared to the number of boxes). In the case shown in figure \ref{Fig:atomsandboxes}, we used 100 boxes. At about 50 atoms ablated, the atomic signal reaches a maximum, and one can see a deviation from a linear behavior down to 10--15 atoms, or 1/8 of the number of boxes So, this model predicts a significant deviation from a linear behavior once about 1/8 of the boxes have been filled.

One can easily get an order of magnitude estimate for the number of boxes in our case of Li in the Ne matrix. We have previously measured~\cite{laserJCP2012} the penetration depth of laser ablated Li atoms reaching a maximum at 6--10 atomic monolayers and extends up to 20--30 monolayers. We take 20 monolayers as typical penetration depth for uniform population. The average distance between Ne atoms in the solid matrix is 2.8~$\AA$. The sapphire area, of order of 8~mm$^2$ would lead to about 10$^{14}$ monolayers squared. Thus, one gets about 2$\times10^{15}$ boxes. This is a typical number of atoms that can be obtained in a single laser ablation pulse in our conditions. Therefore, one can be in a saturated regime (high dilution) even with a single ablation pulse.

The saturation of the atomic and the molecular signals with number of ablations, without the loss of Ne matrix does indicate that some atoms have to be recombining in the matrix. The fact that the numbers fall in the right estimate for saturation also indicates that this process might be happening. On the other hand, clearly our model cannot be considered valid in the regime of very high saturations as the matrix would loose its characteristics (if the impurities reach a level of 50\% one cannot speak of a Ne matrix anymore, it has become another solid). This change in character might be a limiting factor to the density of molecules we can observe, since the molecular signal is a factor of 10--50 lower than our very simple model would predict from the atomic signal (even considering the fact that in some experiments the atoms can interact more than once with the laser beam through absorption and emission cycles, while the molecules, due to the decay to  other states only interact once, we still find a factor of 10 lower molecular signal).

Other questions remain unclear to us: - how does the matrix dissipate the  $\sim$ 8600 cm$^{-1}$  binding energy when atoms recombine? - does that lead to some evaporation of the matrix as we have seen, specially in the case of thin matrices?  - do atoms recombine into molecules if they cross near another in the matrix on their way through the matrix (then a ``cone'' or ``cylinder'' model would describe better than an independent boxes model)? - would an annealing of the matrix help in the molecular formation, to what temperature and for how long?  It would be interesting to see these questions answered by Molecular Dynamics simulations.

Finally, a third possibility by which molecules could be formed in present experimental setup is by atomic combination during the sublimation. If Li atomic density is high, with Ne atoms serving as a third body, Li dimers could be formed. In this case, we expect that they would be initially in high vibrational and rotational states due to the larger phase--space for recombination at high quantum numbers. Thermalization towards low ro-vibrational temperatures, where they would be detectable by the laser, would require a large number of collisions and the translational temperature would be approached from above. In this case, in contrast to what was observed, one would expect a rotational temperature exceeding the translational temperature. We are confident, therefore, that molecules we detect have not been formed during sublimation, but have come as molecules out of Ne solid matrix.

A major concern during the long search for this molecular signal was the possibility that the
molecules could be found in some sort of stretched or compressed configuration within the solid Ne
and, therefore, they could be sent off in several vibrational and rotational states  upon release. In that
case, the narrow laser, tuned to a particular rovibrational state, would not have enough sensitivity to
detect the molecules in an absorption experiment. On the other hand, it may happen that the stretched
or compressed molecules could adiabatically relax to its free configuration during the sublimation
process. Indeed the density is  high during sublimation and the dissipation of this energy could
happen very quickly. In fact, we have not attempted  to detect transitions starting from other vibrational
states away from the ground state ($v''=0$). Thus, we have no upper limit on the vibrational
temperature and cannot rule out the possibility of this process. But, by obtaining a rotational
temperature slightly lower than the translational one, we have an indication that the initial rotational
temperature is low and that this concern may not be happening, or at least it did not prevent us from
observing the ground state molecules. In retrospect, the $^7$Li$_2$ quantum of vibration corresponds to about 500~K (G$_1$ - G$_0$ for the X$^1\Sigma_g^+$ state \cite{Coxon}), much more than the typical Ne--Ne binding energy. Thus, it is expected that the Ne solid should conform to the $^7$Li$_2$ configuration, rather than the opposite. While this is the case for $^7$Li$_2$, it may not be the case for some loosely bound molecules.

\section{Conclusion and Prospects}

In this study we report the first pulsed beam of cryogenic molecules ($^7$Li$_2$ dimers) from Matrix Isolation Sublimation. We measured temperatures as low as 4--6~K, which is understood from a dynamical cooling (higher loss of the faster species) after the full matrix have sublimated. The maximum laser absorption, at 5\% with line width of 0.2 GHz, points to a $^7$Li$_2$ molecular density around 10$^9~{\rm cm}^{-3}$ averaged along the full laser path.  By studying the A$^1\Sigma_u^+(v'=4,J'=J''+1)\leftarrow $X$^1\Sigma_g^+(v''=0,J''=0,1,3)$ transitions we measured a rotational temperature around 4--5~K, slightly lower than the longitudinal translational temperature of 6--8~K for the same series. The translational temperatures are in agreement with previous models \cite{laserJCP2012}, while the forward velocities of the beam seem higher, around $\sim$130~m$\,$s$^{-1}$ for a sublimation temperature of  16~K, for this same set of data. but these values show that it is a competitive technique for performing high precision measurements on ground state molecules.

We have seen, in all the series attempted, a clear saturation of the atomic and the molecular signals which go through a maximum and decay with increasing number of sublimation pulses. By considering a simple model where the Ne matrix serves as a set of boxes where atoms are thrown randomly, and by considering a box with 2 atoms as a formation of a dimer in the matrix, we can mimic this observation. Clearly there is a saturation in the matrix and a detailed balance of the number of implanted atoms with the number of sites in the Ne matrix, together with our previous measurements of the depth of penetration of the lithium atoms, make a compatible picture with observation of this saturation. If indeed a relevant fraction of the molecules are  being formed in the matrix, this may turn out a powerful technique to produce other species of more interest which are difficult to be made in a way compatible to high precision experiments. We are preparing an experiment with heteronuclear molecules to confirm this preliminary evidence that the molecules can be formed in the matrix and to carry on a physics program, in precision tests of physics, quantum information and scientific metrology with these molecules.

While we see no evidence that the molecules would be released in a variety of vibrational states due to a possible stretch or compression in the solid Ne matrix, we cannot rule out completely this possibility for other loosely bound molecules. Observing a rotational temperature lower than the translational is suggestive that the solid Ne matrix is accommodating itself to the ground state $^7$Li$_2$ conformation, rather than the other way around.

%\ack
We acknowledge the work of Douglas Teixeira, Henrique Bergallo in some phases of this study as well as a help from Ivan Silva and Mauricio Lima from INMETRO. This work is partially supported by the Brazilian institutions CNPq, CAPES, and INCT-IQ.

%\section*{References}

\end{document}